\documentstyle[12pt]{article}

\catcode`\@=11
\long\def\@makefntext#1{
\protect\noindent \hbox to 3.2pt {\hskip-.9pt
$^{{\ninerm\@thefnmark}}$\hfil}#1\hfill}                %CAN BE USED

\def\@makefnmark{\hbox to 0pt{$^{\@thefnmark}$\hss}}  %ORIGINAL

\def\ps@myheadings{\let\@mkboth\@gobbletwo
\def\@oddhead{\hbox{}
\rightmark\hfil\ninerm\thepage}
\def\@oddfoot{}\def\@evenhead{\ninerm\thepage\hfil
\leftmark\hbox{}}\def\@evenfoot{}
\def\sectionmark##1{}\def\subsectionmark##1{}}

\renewcommand{\thefootnote}{\fnsymbol{footnote}}

\newcounter{sectionc}\newcounter{subsectionc}\newcounter{subsubsectionc}
\renewcommand{\section}[1] {\vspace*{0.6cm}\addtocounter{sectionc}{1}
\setcounter{subsectionc}{0}\setcounter{subsubsectionc}{0}\noindent
        {\normalsize\bf\thesectionc. #1}\par\vspace*{0.4cm}}
\renewcommand{\subsection}[1] {\vspace*{0.6cm}\addtocounter{subsectionc}{1}
        \setcounter{subsubsectionc}{0}\noindent
        {\normalsize\it\thesectionc.\thesubsectionc. #1}\par\vspace*{0.4cm}}
\renewcommand{\subsubsection}[1]
{\vspace*{0.6cm}\addtocounter{subsubsectionc}{1}
        \noindent
{\normalsize\rm\thesectionc.\thesubsectionc.\thesubsubsectionc.
        #1}\par\vspace*{0.4cm}}

\renewenvironment{thebibliography}[1]
        {\begin{list}{\arabic{enumi}.}
        {\usecounter{enumi}\setlength{\parsep}{0pt}
\setlength{\leftmargin 0.52cm}{\rightmargin 0pt}
         \setlength{\itemsep}{0pt} \settowidth
        {\labelwidth}{#1.}\sloppy}}{\end{list}}

\newcounter{itemlistc}
\newcounter{romanlistc}
\newcounter{alphlistc}
\newcounter{arabiclistc}

\def\@citex[#1]#2{\if@filesw\immediate\write\@auxout
        {\string\citation{#2}}\fi
\def\@citea{}\@cite{\@for\@citeb:=#2\do
        {\@citea\def\@citea{,}\@ifundefined
        {b@\@citeb}{{\bf ?}\@warning
        {Citation `\@citeb' on page \thepage \space undefined}}
        {\csname b@\@citeb\endcsname}}}{#1}}

\newif\if@cghi
\def\cite{\@cghitrue\@ifnextchar [{\@tempswatrue
        \@citex}{\@tempswafalse\@citex[]}}
\def\citelow{\@cghifalse\@ifnextchar [{\@tempswatrue
        \@citex}{\@tempswafalse\@citex[]}}
\def\@cite#1#2{{$\null^{#1}$\if@tempswa\typeout
        {IJCGA warning: optional citation argument
        ignored: `#2'} \fi}}

 1
 1
 1

\font\ninerm=cmr9

\newcommand{\beq}{\begin{equation}}
\newcommand{\eeq}{\end{equation}}
\renewcommand{\a}{\alpha}
\renewcommand{\b}{\beta}
\renewcommand{\d}{\delta}
\newcommand{\eq}[1]{Eq.~(\ref{#1})}
\newcommand{\eqs}[2]{Eq.~(\ref{#1}--\ref{#2})}
\newcommand{\g}{\gamma}
\newcommand{\hh}[2]{{#1}_{t_1},{#1}_{t_2},\ldots,{#1}_{t_#2}}
\newcommand{\ie}{{\em i.e.,\ }}
\newcommand\mathC{\mkern1mu\raise2.2pt\hbox{$\scriptscriptstyle|$}
                {\mkern-7mu\rm C}}
\newcommand{\mathR}{{\rm I\! R}}
\newcommand{\r}{\rho}
\newcommand{\tr}{{\rm tr}}

\newcommand{\D}{{\cal D}}
\newcommand{\UP}{{\cal UP}}
\renewcommand{\H}{{\cal H}}

\begin{document}

% Next two lines used in preprint version only!
\hfill Imperial/TP/94-95/37
\vspace*{1cm}

\centerline{\normalsize\bf QUANTUM LOGIC AND DECOHERING HISTORIES
\footnote{Lecture given at the conference {\em Theories of
Fundamental Interactions\/}, Maynooth, Eire, 24--26 May 1995.}
		}

\vspace*{0.6cm}
\centerline{\footnotesize C.~J.~ISHAM}
\vspace*{0.2cm}
\baselineskip=13pt
\centerline{\footnotesize\em Blackett Laboratory}
\centerline{\footnotesize\em Imperial College of Science, Technology
				\& Medicine}
\centerline{\footnotesize\it South Kensington, London SW7 2BZ, U.K.}
\vspace*{0.1cm}
\centerline{\footnotesize E-mail: c.isham@ic.ac.uk}
\vspace*{0.9cm}

{\centering{\begin{minipage}{12.2truecm}\footnotesize\baselineskip=12pt
\noindent\centerline{\footnotesize ABSTRACT}
\vspace*{0.1cm}
\parindent=0pt

An introduction is given to an algebraic formulation and
generalisation of the consistent histories approach to quantum
theory. The main technical tool in this theory is an orthoalgebra of
history propositions that serves as a generalised temporal analogue
of the lattice of propositions of standard quantum logic. Particular
emphasis is placed on those cases in which the history propositions
can be represented by projection operators in a Hilbert space, and
on the associated concept of a `history group'.
\end{minipage}\par}}
\vspace*{0.6cm}

\normalsize\baselineskip=15pt
\setcounter{footnote}{0}
\renewcommand{\thefootnote}{\alph{footnote}}	% Restore footnote style

\section{Introduction}
In recent years, much attention has been devoted to the so-called
`decoherent histories' approach to quantum theory. A major
motivation for this scheme is a desire to replace the traditional
Copenhagen interpretation of quantum theory with one that avoids any
fundamental split between observer and system and the associated
concept of state-vector reduction induced by a measurement. The key
ingredient of the new approach is an assertion that, under certain
conditions, a probability can be ascribed to a complete {\em
history\/} of a quantum system without invoking any external
state-vector reductions in the development of the history. Any such
scheme would clearly be particularly attractive in quantum cosmology
where a fundamental observer-system split seems to be singularly
inappropriate.

	Whether or not the new approach really does solve the conceptual
challenges of quantum theory has been the subject of much recent
debate; in particular, Dowker and Kent\cite{DK95} have raised some
serious doubts in the context of their penetrating analysis of the
original programme. However, the main concern of the present paper
is not conceptual issues as such but rather the possibility that the
decoherent histories programme could provide a framework for solving
certain technical or structural problems that arise in quantum
gravity.  An example is the `problem of time' that features
prominently in canonical quantum gravity. One plausible conclusion
from the extensive discussion of this issue is that the conventional
notion of time applies only in some semi-classical limit: a
conclusion that, if true, must throw doubt on the entire standard
quantum formalism, depending as it does on certain {\em prima
facie\/} views on the nature of time. One way of tackling this issue
could be with the aid of a suitably generalised notion of a
space-time history.

	However, of even greater importance perhaps is the question of
whether quantum ideas should apply to space-time itself in addition
to the metric or other fields that it carries.  The
inappropriateness of conventional quantum ideas becomes particularly
apparent if one tries to develop non-continuum models of space-time
involving, say, quantised point-set topologies.  As with the
conceptual problems of quantum cosmology, the challenge posed by
issues of this type goes well beyond the question of which
particular approach to quantum gravity (for example: superstring
theory; canonical quantisation) is `correct' by suggesting the need
for a radical reappraisal of quantum theory itself. I believe that a
suitably generalised version of the consistent histories programme
could fulfil this role.

\section{The Main Ideas}
\subsection{The Consistent Histories Formalism in
Normal Quantum Theory}
The consistent histories approach to standard quantum theory was
pioneered by Griffiths\cite{Gri84},
Omn\`es\cite{Omn88a,Omn88b,Omn88c,Omn89,Omn90,Omn92}, and Gell-Mann and Hartle
\cite{GH90a,GH90b,GH90c,Har91a,Har91b,Har95}, and starts
from a result in conventional quantum theory concerning the
joint probability of finding each of a time-ordered sequence of
properties\footnote{A typical property is that the value of some
physical quantity lies in some specified range.}$\;$ $\a=(\hh\a n)$ with
$t_1<t_2<\cdots<t_n$ (we shall call a sequence of this type a {\em
homogeneous history\/}, and refer to the sequence of times as the
{\em temporal support\/} of the history).  Namely, if the initial
state at time $t_0$ is a density matrix $\r_{t_0}$ then the joint
probability of finding all the properties in an appropriate sequence
of measurements is
\beq
	{\rm Prob}(\hh\a n;\r_{t_0})=\tr_\H(C_\a\r_{t_0}C_\a^\dagger)
									\label{Prob:a1-an}
\eeq
where the `class' operator $C_\a$ is given in terms of the
Schr\"odinger-picture projection operators $\a_{t_i}$ on the Hilbert
space $\H$ as
\beq
	C_\a:=U(t_0,t_n)\a_{t_n} U(t_n,t_{n-1})\a_{t_{n-1}}\ldots U(t_2,t_1)
	\a_{t_1}U(t_1,t_0)                              \label{Def:C_a}
\eeq
where $U(t,t')=e^{-i(t-t')H/\hbar}$ is the unitary time-evolution
operator from time $t'$ to $t$. We note
in passing that $C_\a$ is often written as the product of
projection operators
\beq
     C_\a=\a_{t_n}(t_n)\ldots\a_{t_2}(t_2)\a_{t_1}(t_1)\label{Def:C_aH}
\eeq
where $\a_{t_i}(t_i):= U(t_i,t_0)^\dagger\a_{t_i} U(t_i,t_0)$ is
the Heisenberg picture operator defined with respect to the fiducial
time $t_0$.

    The main assumption of the consistent-histories interpretation of
quantum theory is that, under appropriate conditions, the probability
assignment \eq{Prob:a1-an} is still meaningful for a {\em closed\/}
system, with no external observers or associated measurement-induced
state-vector reductions (thus signalling a move from `observables' to
`beables'). The satisfaction or otherwise of these conditions (the
`consistency' of a complete set of histories: see below) is determined
by the behaviour of the {\em decoherence function\/}
$d_{(H,\rho)}$. This is the complex-valued function of pairs of
homogeneous histories $\a=(\hh{\a}n)$ and
$\b=(\b_{t_1'},\b_{t_2'},\ldots,\b_{t_m'})$ defined as
\beq
    d_{(H,\rho)}(\a,\b):=\tr_{\H}(C_\a\rho C_\b^\dagger)  \label{Def:d}
\eeq
where the temporal supports of $\a$ and $\b$ need not be the same.
The physical interpretation of the complex number
$d_{(H,\r)}(\a,\b)$ is as a measure of the extent to which the
histories $\a$ and $\b$ are incompatible in the sense that it is not
meaningful to assert ``either $\a$ is realised {\em or\/} $\b$ is
realised''\footnote{A paradigmatic example of such a situation is
the pair of paths that could be followed classically by a particle
in the two-slit experiment.}$\;$. A key ingredient in the formalism is
the idea of finding collections of projectors that are sufficiently
coarse (\ie project onto sufficiently large subspaces of $\H$) that
the decoherence function of pairs of such can vanish.

	Note that, as suggested by the notation $d_{(H,\rho)}$, both the
initial state and the dynamical structure (\ie the Hamiltonian $H$)
are coded in the decoherence function. A homogeneous history
$(\hh{\a}n)$ itself is just a `passive', time-ordered sequence
of propositions that can be read as the single sequential
proposition ``$\a_{t_1}$ is true at time $t_1$, and then $\a_{t_2}$
is true at time $t_2$, and then \ldots, and then $\a_{t_n}$ is true
at time $t_n$''.

\subsection{Generalised History Theory}
An important suggestion of Gell-Mann and Hartle was to develop a new
type of quantum theory in which the ideas of `history' and
`decoherence function' would be fundamental in their own right. In
particular, a history need no longer be just a time-ordered sequence
of projection operators. They suggested that the crucial ingredients
in such a theory would be (i) a `coarse-graining' operation on the
generalised histories; (ii) a mechanism for forming a logical `or'
of a pair of `disjoint' histories (so that, in certain
circumstances, one can talk about ``history $A$ or history $B$''
being realised); and (iii) a negation operation (so that, in
appropriate circumstances, one can make assertions like ``history
$A$ is {\em not\/} realised'').

	Much of their thinking on this matter was motivated by path
integrals where a typical coarse-grained history is that the path in
the configuration space $Q$ lies in some specified subset of paths.
Thus they defined the decoherence function
\beq
    d(\a,\b):=\int_{q\in\a,q'\in\b}Dq\,Dq'\,e^{-i(S[q]-S[q'])/\hbar}
            \d\big(q(t_1),q'(t_1)\big)\rho\big((q(t_0),q'(t_0)\big)
                            \label{Def:dPI}
 \eeq
where the integral is over paths that start at time $t_0$ and end at
time $t_1$, and where $\a$ and $\b$ are subsets of paths in $Q$. In
this case, to say that a pair of histories $\a$ and $\b$ is disjoint
means simply that they are disjoint subsets of the path space of $Q$,
in which case $d$ clearly possesses the additivity property
\beq
        d(\a\oplus\b,\g)=d(\a,\g)+d(\b,\g)
\eeq
for all subsets $\g$ of the path space. Similarly, $\neg\a$ is
represented by the complement of the subset $\a$ of path space, in
which case the decoherence function satisfies
\beq
        d(\neg\a,\g)= d(1,\g)-d(\a,\g)
\eeq
where $1$ denotes the entire path space (the `unit' history).

\subsection{An Algebraic Scheme}
I would now like to summarise the algebraic scheme proposed by Isham
and Linden\cite{Ish94,IL94} for placing the Gell-Mann and Hartle
scheme in a precise mathematical framework that brings out the
natural relation to concepts in quantum logic. In studying this
rather abstract scheme it is appropriate to keep in mind that, in
practice, the notion of a generalised history can include many
different types of mathematical object. For example, in the case of
quantum gravity, a `history' could be a (possibly non
globally-hyperbolic) space-time geometry (or subset of such), or a
geometry augmented with other fields. Or it could include a
specification of the space-time manifold---thereby describing a type
of quantum topology---or the `history' could even be an arbitrary
topological space. Of particular importance is the idea that the
class of generalised histories will generically include
`non-abelian' versions of the above. By this is meant some analogue
of the fact that, in a class operator like \eq{Def:C_a}
in standard quantum theory, the Schr\"odinger-picture projectors
$\a_{t_i}$ at different times $t_i$ may not commute (for example,
they could include both position and momentum projectors), unlike
the projectors onto subspaces of configuration space that arise in
normal path integrals. Indeed, in some cases, the ideas that follow
can be viewed as defining a non-commutative version of a path
integral.
\smallskip

	The basic rules of our version of the Gell-Mann and Hartle
axioms are as follows\cite{Ish94,IL94}.

\begin{enumerate}
\item The fundamental ingredients in the theory are (i) a space
$\UP$ of propositions about possible `histories' (or `universes');
and (ii) a space $\D$ of {\em decoherence functions\/}. A
decoherence function is a complex-valued\footnote{This is the
precise point at which complex numbers enter the generalised scheme.
Complex numbers are used in analogy to what occurs in standard
quantum theory in the context of the class function \eq{Def:C_a} or
the path integral \eq{Def:dPI}. However, this does not rule out
other possibilities for the space in which the decoherence functions
take their values.}$\;$ function of pairs $\a,\b\in\UP$ whose value
$d(\a,\b)$ is a measure of the extent to which the history
propositions $\a$ and $\b$ are `mutually incompatible'.  The pair
$(\UP,\D)$ is to be regarded as the generalised-history analogue of
the pair $({\cal L},{\cal S})$ in standard quantum theory where
$\cal L$ is the set of propositions about the system at some fixed
time, and $\cal S$ is the space of quantum states.

 \item{The set $\UP$ of history propositions is equipped with the
following, logical-type, algebraic operations:

	\begin{enumerate}
		\item{A partial order $\leq\,$. If $\a\leq\b$ then $\b$ is
said to be {\em coarser\/} than $\a$, or a {\em coarse-graining\/}
of $\a$; equivalently, $\a$ is {\em finer\/} than $\b$, or a {\em
fine-graining\/} of $\b$. The heuristic meaning of this relation is
that $\a$ provides a more precise affirmation of `the way the
universe is' (in a transtemporal sense) than does $\b$.

    The set $\UP$ possesses a {\em unit\/} history proposition
$1$---heuristically, the proposition about possible
histories/universes that is always true---and a {\em null
history proposition\/} $0$---heuristically, the proposition that is
always false. For all $\a\in\UP$ we have $0\leq\a\leq 1$.
			}

    	\item{There is a notion of two history propositions $\a,\b$ being
{\em disjoint\/}, written $\a\perp\b$. Heuristically, if $\a\perp\b$
then if either $\a$ or $\b$  is `realised' the other certainly cannot be.

    Two disjoint history propositions $\a,\b$ can be combined to
form a new proposition $\a\oplus\b$ which, heuristically, is the
proposition `$\a$ {\em or\/} $\b$'. This partial\footnote{It should
be noted that the structure of an orthoalgebra is much weaker than
that of a lattice. In the latter, there are two connectives $\land$
and $\lor$, both of which are defined on {\em all\/} pairs of
elements. This contrasts with the single, partial operation $\oplus$
in an orthoalgebra. A lattice is a special type of orthoalgebra,
with $a\oplus b$ being defined on disjoint lattice elements $a,b$ as
$a\lor b$. Here, `disjoint' means that $a\leq\neg b$.}$\;$ binary
operation is assumed to be commutative and associative, \ie
$\a\oplus\b=\b\oplus\a$, and $\a\oplus(\b\oplus\g)=
(\a\oplus\b)\oplus\g$ whenever these expressions are meaningful.  }

    	\item There is a negation operation $\neg\a$ such that, for all
$\a\in\UP$, $\neg(\neg\a)=\a$.

		\end{enumerate}
	}
\end{enumerate}

    A crucial question is how the operations $\leq$, $\oplus$ and
$\neg$ are to be related. We shall postulate the following, minimal,
requirements:\footnote{Note that the second condition is manifestly
true for a Boolean algebra (in which case, without loss of
generality, the $\leq$ ordering can be regarded as set inclusion)$\;$,
and also for the algebra of projection operators on a Hilbert space,
in which $P\leq Q$ means that $P$ projects onto a subspace of the
range of $Q$ (one then writes $Q$ as the sum of $P$ and the
projector onto the orthogonal complement of the range of $P$ in the
range of $Q$).}
\begin{eqnarray*}
    i)&& \neg\a
        {\rm\ is\ the\ unique\ element\ in\ }\UP{\rm\ such\ that\ }
        \a\perp\neg\a {\rm\ with\ } \a\oplus\neg\a=1;      \\
    ii)&& \a\leq\b {\rm\ if\ and\ only\ if\ there\ exists\ } \g\in\UP
        {\rm \ such\ that\ } \b=\a\oplus\g.
\end{eqnarray*}

	 Both conditions are true of, for example, subsets of paths in a
configuration space and, together with the other requirements
above, essentially say that $\UP$ is an {\em orthoalgebra\/}; for a
full definition see Foulis {\em et al\/}\cite{FGR92}. One
consequence is that
\beq
        \a\perp\b {\rm\ if\ and\ only\ if\ }\a\leq\neg\b.
\eeq
An orthoalgebra is probably the minimal useful mathematical
structure that can be placed on $\UP$, but of course that does not
prohibit the occurrence of a stronger one; in particular $\UP$ could
be a lattice\footnote{One virtue of the weaker structure is that no
one has been able to define a satisfactory tensor product for
lattices whereas this {\em is\/} possible for orthoalgebras
\cite{FGR92}.}$\;$. The possibility of generalising the structure of
$\UP$ to be that of a `difference poset' has been suggested recently
by Pulmanova\cite{Pul95}. This allows the propositions to be
extended to include `effects': a possibility that is of some
importance in quantum theory in general\cite{BLM91}.

	The next step is to formalise the notion of a decoherence
function. Specifically, a {\em decoherence function\/} is a map
$d:\UP\times\UP\rightarrow\mathC$ that satisfies the following
conditions:
\begin{enumerate}
	\item {\em Hermiticity\/}: $d(\a,\b)=d(\b,\a)^*$ for all $\a,\b$.
	\item {\em Positivity\/}: $d(\a,\a)\ge0$ for all $\a$.
	\item {\em Additivity\/}: if $\a\perp\b$ then, for all $\g$,
		$d(\a\oplus\b,\g)=d(\a,\g)+d(\b,\g)$. If appropriate, this can be
		extended to countable sums.
	\item {\em Normalisation\/}: $d(1,1)=1$.
\end{enumerate}
\noindent
In addition to the above we adopt the following definitions of
Gell-Mann and Hartle: A set of history propositions
$\{\a^1,\a^2,\ldots,\a^N\}$ is said to be {\em exclusive\/} if
$\a^i\perp\a^j$ for all $i,j=1,2,\ldots,N$. The set is {\em
exhaustive\/} (or {\em complete\/}) if it is exclusive and if
$\a^1\oplus\a^2\oplus\ldots\oplus\a^N=1$. In algebraic terms, an
exclusive and exhaustive set of history propositions is simply a
{\em partition of unity\/} in the orthoalgebra $\UP$.

    It must be emphasised that, within this scheme, only {\em
consistent\/} sets of history propositions are given an immediate
physical interpretation. A complete set $\cal C$ of history
propositions is said to be (strongly) consistent with respect to a
particular decoherence function $d$ if $d(\a,\b)=0$ for all
$\a,\b\in{\cal C}$ such that $\a\ne\b$. Under these circumstances
$d(\a,\a)$ is regarded as the {\em probability\/} that the history
proposition $\a$ is true.  The axioms above then guarantee that the
usual Kolmogoroff probability rules are satisfied on the Boolean
algebra generated by $\cal C$.

	In this context, it is worth remarking that the idea of an
orthoalgebra is closely related to that of a {\em Boolean
manifold\/}\footnote{In turn, this is closely related to the idea of
a {\em manual\/}: a concept that has been developed extensively in
standard quantum logic by Foulis and Randall (see Foulis {\em et
al\/}\cite{FGR93} and references therein). In many respects this
structure seems the most appropriate of all in which to develop a
generalised history theory; however, this remains a task for the
future.}$\;$: an algebra that is `covered' by a collection of
Boolean subalgebras with appropriate compatibility conditions on any
pair that overlap\cite{HF81}. Being Boolean, these subalgebras of
propositions carry a logical structure that is essentially
classical: a feature of the decoherent histories scheme that was
particularly emphasised in the seminal work of Griffiths and
Omn\`es. In the approach outlined above, these Boolean algebras are
glued together from the outset to form a universal algebra $\UP$ of
propositions from which the physically interpretable subsets are
selected by the consistency conditions with respect to a chosen
decoherence function.

	For a different perspective that places the emphasis on the
separate Boolean algebras see the recent paper by
Griffiths\cite{Gri95} in which he emphasises the dangers that can
occur if logical deductions arising from incompatible consistent
sets are mixed together. Dangers of this type are potentially
present in all uses of quantum logic, and one must be very careful
not to assume {\em a priori\/} that the algebraic operations
employed have a logical interpretation in any semantic sense even
though, mathematically speaking, they do look like logical (albeit,
non-distributive) connectives.

\section{Some Key Results}
I will now summarise some of the main results that have been
achieved in this quantum-logic like approach to the generalised
histories programme. For further details the reader should
consult the original papers\cite{Ish94,IL94,ILS94,IL95}. It
should be noted that the results discussed here all concern the
important special case in which the history propositions can be
represented by projectors on some Hilbert space.

\subsection{Inclusion of Standard Quantum Theory in the Scheme}
The similarity of the axioms above to those of conventional quantum
logic motivates investigating the possibility of representing
history propositions with projection operators on a Hilbert space.
In particular, the question arises if it is possible to find such a
representation for the homogeneous history propositions of standard
quantum theory. Note that this is not a trivial matter since the
product of two projection operators $P$ and $Q$ (such as appears in
the class operator \eq{Def:C_aH}) is not itself a projector unless
$[\,P,Q\,]=0$.

	The key to resolving this issue is the observation that what we
are seeking is a quantum version of {\em temporal\/} logic rather
than the logic of single-time propositions used in most discussions
of physics. To this end, consider a temporal-logic sequential
conjunction $A\sqcap B$ to be read as ``$A$ is true and then $B$ is
true''. Then the proposition $A\sqcap B$ is false if (i) $A$ is
false and then $B$ is true, or (ii) $A$ is true and then $B$ is
false, or (iii) $A$ is false and then $B$ is false; symbolically:
\beq
    \neg(A\sqcap B)=\neg A\sqcap B\ or\ A\sqcap\neg B\
                or\ \neg A\sqcap\neg B.                 \label{negAthenB}
\eeq

	Now we make the crucial observation that, unlike the simple
product $PQ$, the {\em tensor\/} product $P\otimes Q$ of a pair of
projection operators $P,Q$ on a Hilbert space $\H$ is {\em always\/}
a projection operator. Indeed, the product of homogeneous operators
on $\H\otimes\H$ is defined as $({A}\otimes{B})({C}\otimes {D}):=
{A}{C}\otimes{B}{D}$, while the adjoint operation is
$({A}\otimes{B})^\dagger:={A}^\dagger\otimes{B}^\dagger$, and hence
$(P\otimes Q)^2=P^2\otimes Q^2=P\otimes Q$, and $(P\otimes
Q)^\dagger =P^\dagger\otimes Q^\dagger=P\otimes Q$.

	Since $P\otimes Q$ is a genuine projection operator we
have\footnote{In standard quantum logic, the projector that
represents the negation of a proposition $R$ is $1-R$, which
projects onto the orthogonal complement of the range of $R$.}$\;$ the
relation $\neg(P\otimes Q)= 1\otimes1-P\otimes Q$ on $\H\otimes\H$, and so
\begin{eqnarray}
	\neg(P\otimes Q)=1\otimes1-P\otimes Q&=& (1-P)\otimes Q+P\otimes(1-Q)+
		(1-P)\otimes(1-Q) 							\nonumber \\
	&=& \neg P\otimes Q+P\otimes\neg Q+\neg P\otimes\neg Q
\end{eqnarray}
which exactly models \eq{negAthenB}. This suggests representing the
two-time sequential conjunction ``$\a_{t_1}$ at time $t_1$ and then
$\a_{t_2}$ at time $t_2$'' with the tensor product\footnote{This
representation works well in capturing the essential nature of
temporal propositions. However, it constitutes a striking departure
from conventional thinking about the role of time in quantum theory,
and therefore the idea needs to be handled carefully. For example,
note that even if $\a_{t_1}$ and $\a_{t_2}$ are a pair of
propositions that do {\em not\/} commute, the homogeneous history
projectors $\a_{t_1}\otimes 1_{t_2}$ and $1_{t_1}\otimes\a_{t_2}$
{\em do\/} commute by virtue of the law of tensor product
multiplication.}$\;$ $\a_{t_1}\otimes\a_{t_2}$. Of course, not every
projection operator in\footnote{Both $\H_{t_1}$ and $\H_{t_2}$ are
isomorphic copies of the Hilbert space $\H$ on which the original
quantum theory is defined: the $t_1$ and $t_2$ subscripts in
$\H_{t_1}$ and $\H_{t_2}$ serve only as a reminder of the times to
which the propositions $\a_{t_1}$ and $\a_{t_2}$ refer.}$\;$
$\H_{t_1}\otimes\H_{t_2}$ is of this homogeneous form.  In
particular, an inhomogeneous projection operator like
$\a_{t_1}\otimes\a_{t_2}+\b_{t_1}\otimes\b_{t_2}$ can represent the
proposition ``($\a_{t_1}$ at time $t_1$ and then $\a_{t_2}$ at time
$t_2$) {\em or} ($\b_{t_1}$ at time $t_1$ and then $\b_{t_2}$ at
time $t_2$)'' provided that the projectors $\a_{t_1}\otimes\a_{t_2}$
and $\b_{t_1}\otimes\b_{t_2}$ are disjoint\footnote{In general, a
pair of projectors $P$ and $Q$ is {\em disjoint\/} if $PQ=0$.}$\;$.
History propositions of this type (\ie sums of disjoint homogeneous
history propositions) are called {\em inhomogeneous\/} and are an
important generalisation of the idea of a history proposition.

	Further investigation shows that this idea of using tensor
products works very well and, in general, the homogeneous $n$-time
history proposition $\hh \a n$ can be represented by the projection
operator $\a_{t_1}\otimes\a_{t_2}\otimes \cdots\otimes\a_{t_n}$ on
the tensor product
$\H_{t_1}\otimes\H_{t_2}\otimes\cdots\otimes\H_{t_n}$.
Inhomogeneous history propositions then correspond to sums of
pairwise-disjoint homogeneous history propositions. By this means
the theory of discrete-time histories in standard quantum theory can
be placed in the framework of the axiomatic scheme above.  In
particular, it can be shown that the decoherence function
$d_{(H,\r)}(\a,\b)$ in
\eq{Def:d} of a pair of homogeneous history propositions $\a=(\hh\a n)$ and
$\b=(\b_{t_1'},\b_{t_2'},\ldots,\b_{t_m'})$ can be written in terms
of the associated tensor product operators as
\beq
	d_{(H,\r)}(\a,\b)=\tr_{\otimes^{n+m}\H}
		((\a_{t_1}\otimes\a_{t_2}\otimes\cdots\otimes\a_{t_n})\otimes
			(\b_{t_1'}\otimes\b_{t_2'}\otimes\cdots\otimes\b_{t_m'})X)
												\label{Def:dX}
\eeq
for a certain operator $X$ on $\otimes^{n+m}\H$.

\subsection{The Analogue of Gleason's Theorem}
Not all projection operators in a tensor product
$\H_{t_1}\otimes\H_{t_2}\otimes\cdots\otimes\H_{t_n}$ can be written
as sums of {\em disjoint\/} homogeneous histories\footnote{Linden
and I suspect that {\em any\/} projection operator can be obtained
from the set of homogeneous history propositions by the application
of the full {\em lattice\/} operations in the space of projectors,
but we do not know a general proof of this. However, even if true,
it is not clear what the physical significance of this would be.
History propositions that are neither homogeneous or inhomogeneous
have been referred to\cite{IL94} as {\em exotic\/}; an example is
the proposition corresponding to the projector onto an inhomogeneous
vector $u_{t_1}\otimes u_{t_2}+v_{t_1}\otimes v_{t_2}$ in
$\H_{t_1}\otimes\H_{t_2}$. It remains an intriguing topic for
research to see if they have any role to play in our history version
of standard quantum theory.}$\,$.  Nevertheless, the discussion
above raises the question of whether generalised history theories
exist in which the orthoalgebra $\UP$ is the set ${\cal P}({\cal
V})$ of projection operators on some Hilbert space $\cal V$ that is
{\em not\/} just the tensor product of temporally-labelled copies of
a single Hilbert space $\H$.  Indeed, such examples could provide an
extensive source of specific realisations of the axioms. However, to
be viable, this suggestion requires a classification of the possible
decoherence functions on ${\cal P}({\cal V})$: a problem that is a
direct analogue of that solved by Gleason\cite{Gle57} in his famous
theorem in standard quantum logic.

	In standard quantum theory on a Hilbert space $\H$, a state is
defined to be a function $\sigma:{\cal P}(\H)\rightarrow\mathR$ with
the following properties:
\begin{enumerate}
	\item {\em Positivity\/}: $\sigma(P)\ge0$ for all $P\in{\cal P}(\H)$.

    \item {\em Additivity\/}: if $P$ and $R$ are disjoint projectors then
          $\sigma(P\oplus R)=\sigma(P)+\sigma(R)$. This requirement
          is usually extended to include countable collections of
          propositions.

    \item {\em Normalisation\/}: $\sigma(1)=1$
\end{enumerate}
where the unit operator $1$ on the left hand side represents the
unit proposition that is always true. Gleason's theorem asserts
that, when $\dim\H>2$, such states are in one-to-one
correspondence with density matrices $\r$ on $\H$, with
\beq
		\sigma(P)=\tr(P\r)					\label{Gleason}
\eeq
for all projection operators $P\in{\cal P}(\H)$.

	The analogous result for decoherence functions was proved
by Isham, Linden and Schreckenberg\cite{ILS94}. Specifically, the
decoherence functions $d\in\D$ on a space of projectors\footnote{\ In
the original proof only finite-dimensional Hilbert spaces $\cal V$
were discussed.} $\ {\cal P}({\cal V})$ (with $\dim{\cal V}>2$) are in
one-to-one correspondence with operators $X$ on the tensor product
${\cal V}\otimes{\cal V}$ with
\beq
   d_X(\a,\b)=\tr(\a\otimes\b X){\rm\ for\ all\ }\a,\b\in{\cal P}({\cal V}),
                                                        \label{main}
\eeq
and where $X$ satisfies:
\begin{enumerate}
    \item $X^\dagger=MXM$ where the operator $M$ is defined on
          ${\cal V}\otimes{\cal V}$ by $M(u\otimes v):=v\otimes u$;

    \item for all $\a\in{\cal P}({\cal V})$, $\tr(\a\otimes\a\,
	X_1)\geq0$ where $X=X_1+iX_2$ with $X_1$ and $X_2$ hermitian;

    \item $\tr(X_1)=1$.
\end{enumerate}
Note that \eq{Def:dX} is the particular form taken by this expression in the
case of standard quantum theory.

	A simple illustration of this theorem has been given by
Schreckenberg\cite{Sch95}.  Let $\{\a^1,\a^2,\ldots,\a^N\}$ be any
partition of unity in ${\cal P}({\cal V})$, so that
$\a^1+\a^2+\cdots +\a^N=1$ and $\a^i\a^j=\d^{ij}\a^i$.  For any
weights $w_1,w_2,\ldots,w_N$ with $w_i>0$ and $\sum_{i=1}^Nw_i=1$,
define $X:=\sum_{i=1}^Nw_i\,\a_i\otimes\a_i$. Then it is easy to see
that the history propositions represented by
$\{\a^1,\a^2,\ldots,\a^N\}$ are a consistent set with respect to the
decoherence function $d_X$ defined by this particular choice of $X$.

	It should be noted that whatever analogue there may be of both
dynamics and initial conditions is coded into the structure of the
single operator $X$. In the example \eq{Def:dX} that pertains to standard
Hamiltonian quantum theory the operator $X$ takes on a very special
form. However, the theorem stated above for the general case
where $\UP={\cal P}({\cal V})$ for some $\cal V$ allows for a wide
range of possible operators $X$ and hence for a wide range of
generalisations of dynamics and initial conditions. This is the
basis of our hope that the generalised scheme may provide a powerful
tool for handling physical situations in which the notion of time is
non-standard, such as that arising in canonical quantum gravity or
in more exotic programmes aiming at quantising the stucture of
space-time itself.

	The proof of the classification of decoherence functions uses
Gleason's theorem which, over the years, has been generalised to a
variety of types of algebraic structure. Not surprisingly, a similar
situation holds for decoherence functions and, in particular,
Wright\cite{Wri95} has recently extended the classification theorem
to the case where the history propositions are represented by
projections in an arbitrary\footnote{\ Strictly speaking, the
von~Neumann algebra has to have no direct summand of type $I_2$.}$\;$
von~Neumann algebra. The basis of his work is an earlier
result\cite{BMW94} detailing the conditions under which a state
defined on the projectors in a von~Neumann algebra $\cal A$ can be
extend to a linear functional on the entire algebra.  When applied
to the history situation this result can be used to show that a
bounded decoherence functional can be extended to a bilinear function on
$\cal A$; a Gelfand--Naimark--Segal type of construction (which is
naturally suggested by the `inner product' nature of the defining
conditions of a decoherence functional) then completes the process.

\subsection{The History Group}
In standard canonical quantum theory an important role in
constructing specific theories is played by the group of canonical
commutation relations. For example, the quantum theory of a point
particle moving in one dimension is specified by requiring the
Hilbert space to carry an irreducible representation of the
Weyl--Heisenberg group $\cal W$ whose Lie algebra is associated with
the familiar commutation relation $[\,x,p\,]=i\hbar$. The famous
Stone and von~Neumann theorem then shows that the familiar
representation on wave functions is essentially unique. More
generally, if the classical configuration space is a homogeneous
space $G/H$ then the quantisation can be associated\cite{Ish84} with
irreducible representations of a new canonical group constructed
from $G$.

	The question of interest is whether there may be an analogue of
the canonical group in a history theory whose propositions are
associated with projectors on some Hilbert space $\cal V$ as
discussed above. To explore this issue let us question again the
origin of the representation of a homogeneous history proposition
(with temporal support $\{t_1,t_2,\ldots,t_n\}$) in standard quantum
theory by a projection operator on the tensor product
$\H_{t_1}\otimes\H_{t_2}\otimes\cdots\otimes\H_{t_n}$ of $n$ copies
of the original Hilbert space $\H$.

	One answer is the temporal-logic approach that was
sketched earlier. Another option is to invoke the purely algebraic
fact\cite{ILS94} that the trace of a product of operators
$A_1,A_2,\ldots, A_n$ on a Hilbert space $\H$ can be written as a
trace on $\otimes^n\H$ in the form
\beq
	\tr_{\H}(A_1A_2\ldots A_n)\equiv\tr_{\otimes^n\H}(A_1\otimes
		A_2\otimes\cdots\otimes A_n\,S)
\eeq
for a certain fixed operator $S$ on $\otimes^n\H$.

	However, in the case where $\H=L^2(\mathR)$ (\ie the Hilbert
space of wave functions used in elementary wave mechanics) one could
also say that the history Hilbert space $L^2_{t_1}(\mathR)\otimes
L^2_{t_2}(\mathR)\otimes\cdots\otimes L^2_{t_n}(\mathR)$ comes from
a representation of the product $W_{t_1}\times
W_{t_2}\times\cdots\times W_{t_n}$ of $n$ copies of the
Weyl--Heisenberg group, one for each time slot in the temporal
support of the history proposition. Thus the tensor-product Hilbert
space could be viewed as arising as a representation of the
`temporally-gauged' (!) canonical algebra\footnote{I am are assuming
here that the value of $\hbar$ is the same at each time slot in the
temporal support.}
\begin{eqnarray}
		{[}\,x_{t_i},x_{t_j}\,{]}&=&0				\label{xixj}	\\
		{[}\,p_{t_i},p_{t_j}\,{]}&=&0				\label{pipj}	\\
		{[}\,x_{t_i},p_{t_j}\,{]}&=&i\hbar\d_{ij}.	\label{xipj}
\end{eqnarray}

	This observation motivates the intriguing idea that it may be
possible to specify generalised history theories by finding an
appropriate `history group' $\cal G$ whose irreducible unitary
representations give the Hilbert space $\cal V$ on which the history
propositions are to be defined.  In particular, the projectors in
the spectral decompositions of the self-adjoint generators of $\cal
G$ will give a preferred class of propositions---rather as the
generators of a standard canonical group provide a special class of
classical observables that can be represented unambiguously in the
quantum theory\cite{Ish84}.

\subsection{Continuous Histories}
The use of a history group has been illustrated recently by Isham
and Linden\cite{IL95} in the context of {\em continuous\/} time
histories in standard quantum theory. An obvious problem when handling
continuous histories is to define an appropriate continuous
product $\prod_{t\in\mathR}\a_t$ of projection operators for use in a
class operator. However, we have shown\cite{IL95} that this can be
done for projections onto coherent states, and explicit expressions
have been given for this product as well as for the associated
decoherence function of a pair of such continuous histories.

	On the other hand, the discussion above of a history group suggests
that, in the case of continuous histories, the appropriate analogue
of \eqs{xixj}{xipj} is
\begin{eqnarray}
			{[}\,x_t,x_{t'}\,{]}&=& 0					\\
			{[}\,p_t,p_{t'}\,{]}&=& 0					\\
			{[}\,x_{t},p_{t'}\,{]}&=&i\hbar\d(t-t').
\end{eqnarray}
Thus the continuous-time history version of one-dimensional wave
mechanics looks like a one-dimensional quantum field theory, but
with the `fields' being labelled\footnote{It is most important {\em
not\/} to confuse the time-labelled operators with
Heisenberg-picture operators: the one-parameter families of
operators $x_t$ and $p_t$, $t\in\mathR$, are in the Schr\"odinger
picture.}$\;$ by time rather than space!  In particular, as shown in
Isham and Linden\cite{IL95}, this history-group algebra does indeed
provide the correct Hilbert space for the history theory. A key
technical ingredient is the fact that Bosonic Fock space can be
written as a certain continuous tensor product\cite{Gui72}, thereby
linking the representation of the history group with the idea of
continuous temporal logic.

\section{Conclusion}
We have seen that the generalised history scheme proposed by
Gell-Mann and Hartle can be given a precise mathematical form in
which the roles of the space $\UP$ of history propositions and the
space $\cal D$ of decoherence functions are analogous to those in
standard quantum theory of the space $\cal L$ of single-time
propositions and the space $\cal S$ of states respectively. We saw
that finite-time history propositions in standard quantum theory can
be fitted into this generalised algebraic framework by identifying a
homogeneous history proposition $(\hh\a n)$ with the projection
operator $\a_{t_1}\otimes\a_{t_2}\otimes\cdots\otimes\a_{t_n}$ on
the tensor product
$\H_{t_1}\otimes\H_{t_2}\otimes\cdots\otimes\H_{t_n}$.  The
tensor-product space also provides a natural home for inhomogeneous
history propositions via the disjoint `or' operation defined on
pairs of disjoint homogeneous histories. By this means we arrive at
a concrete implementation of the idea of temporal quantum logic.

	This result suggests that a large number of generalised history
theories might be found by looking at a more general situation in
which history propositions are represented by projectors on some
`history' Hilbert space $\cal V$ that is not necessarily a
temporally-labelled tensor product. This lead us to consider the
analogue of Gleason's theorem for decoherence functions, and hence
to the representation of any such in the form $d_X(\a,\b)=\tr_{{\cal
V}\otimes{\cal V}}(\a\otimes\b X)$.

	Finally, we suggested that the Hilbert space $\cal V$ that
carries the generalised history propositions could arise as an
irreducible representation of a history group $\cal G$---a history
analogue of the canonical group of conventional quantum theory.
Note that, in the context of continuous-time, standard quantum
theory, paths in configuration space correspond to a certain maximal
Boolean subalgebra of the orthoalgebra of history projectors.  Thus
the history group serves to embed this Boolean algebra in a specific
non-Boolean orthoalgebra.  It is in this sense that a decoherence
function can sometimes be understood as a non-commutative analogue
of a standard path integral.

	Generalised history theories of the type discussed above offer a
wide-ranging extension of standard ideas in quantum theory and are
well suited for implementing some of the more exotic ideas often
discussed in the context of quantum gravity. For example, it becomes
quite feasible to consider a scheme in which the basic history
propositions include assertions that the space-time topology belongs
to some particular subset of point-set topologies on a fixed or
variable set of space-time points. A less exotic example would be to
study decoherence functionals and space-time metric propositions
that are manifestly invariant under the action of the space-time
diffeomorphism group. This would be a natural way of using the
quantum history programme to find a space-time oriented approach to
quantum gravity.  Discussions of this and other applications will
appear later.

\section{Acknowledgements}
Most of the work reported here was done with Noah Linden, and I
thank him warmly for his friendship and collaboration over the years. I
would also like to thank the organisers of the Maynooth meeting for
their efforts on behalf of the participants. Partial financial
support was provided by the EEC Network `Physical and Mathematical
Aspects of Fundamental Interactions'.


\section{References}

\begin{thebibliography}{10}
\bibitem{DK95} F.~Dowker and A.~Kent, ``On the consistent histories
	approach to quantum theory'' to appear {\em Jour.\ Stat,\ Phys.}
	(1995), gr-qc/9412067

\bibitem{Gri84} R.~B.~Griffiths, ``Consistent histories and the
	interpretation of quantum mechanics'' {\em Jour.\ Stat.\ Phys.}
	{\bf 36} (1984) 219--272.

\bibitem{Omn88a} R.~Omn\`es, ``Logical reformulation of quantum
	mechanics. I. Foundations'' {\em Jour.\ Stat.\ Phys.} {\bf 53}
	(1988) 893--932.

\bibitem{Omn88b} R.~Omn\`es, ``Logical reformulation of quantum
	mechanics. II. Inferences and the Einstein--Podolsky--Rosen
	experiment'' {\em Jour.\ Stat.\ Phys.} {\bf 53} (1988) 933--955.

\bibitem{Omn88c} R.~Omn\`es, ``Logical reformulation of quantum
	mechanics. III. Classical limit and irreversibility'' {\em
	Jour.\ Stat.\ Phys.} {\bf 53} (1988) 957--975.

\bibitem{Omn89} R.~Omn\`es, ``Logical reformulation of quantum
	mechanics. IV. Projectors in semiclassical physics'' {\em Jour.\
	Stat.\ Phys.} {\bf 57} (1989) 357--382.

\bibitem{Omn90} R.~Omn\`es, ``From Hilbert space to common sense: A
	synthesis of recent progress in the interpretation of quantum
	mechanics'' {\em Ann.\ Phys.}~(N.Y.) {\bf 201} (1990) 354--447.

\bibitem{Omn92} R.~Omn\`es, ``Consistent interpretations of quantum
	mechanics'' {\em Rev.\ Mod.\ Phys.} {\bf 64} (1992) 339--382.

\bibitem{GH90a} M.~Gell-Mann and J.~Hartle, ``Quantum mechanics in
	the light of quantum cosmology'' in {\em Proceedings of the
	Third International Symposium on the Foundations of Quantum
	Mechanics in the Light of New Technology\/}, (Physical Society of
	Japan, Tokyo, Japan, 1990), pp321--343.

\bibitem{GH90b} M.~Gell-Mann and J.~Hartle, ``Quantum mechanics in
	the light of quantum cosmology'' in {\em Complexity, Entropy and
	the Physics of Information, SFI Studies in the Science of
	Complexity, Vol.~VIII}, (Addison-Wesley, Reading, 1990), pp425--458.

\bibitem{GH90c}  M.~Gell-Mann and J.~Hartle, ``Alternative
	decohering histories in quantum mechanics'' in {\em Proceedings
	of the 25th International Conference on High Energy Physics\/},
	(World Scientific, Singapore, 1990).

\bibitem{Har91a} J.~Hartle, ``The quantum mechanics of cosmology''
	in {\em Quantum Cosmology and Baby Universes\/}, eds S.~Coleman,
	J.~Hartle, T.~Piran and S.~Weinberg (World Scientific, Singapore,
	1991).

\bibitem{Har91b} J.~Hartle, ``Space-time grainings in
	nonrelativistic quantum mechanics'' {\em Phys.\ Rev.} {\bf D44}
	(1991) 3173--3195.

\bibitem{Har95} J.~Hartle, ``Space-time quantum mechanics and the
	quantum mechanics of space-time'' in {\em Gravitation and
	Quantizations: Les Houches 1992--Session LVII}, eds B.~Julia and
	J.~Zinn-Justin (Elsevier Science, Netherlands, 1995), pp1--62.

\bibitem{Ish94} C.~J.~Isham, ``Quantum logic and the histories
	approach to quantum theory'' {\em Jour.\ Math.\ Phys.} {\bf 23}
	(1994) 2157--2185.

\bibitem{IL94} C.~J.~Isham and N.~Linden, ``Quantum temporal logic
	and decoherence functionals in the histories approach to generalised
	quantum theory'' {\em Jour.\ Math.\ Phys.} {\bf 35} (1994)
	5452--5476.

\bibitem{FGR92} D.~J.~Foulis, R.~J.~Greechie and G.~T.~R\"uttiman,
	``Filters and supports in orthoalgebras'' {\em Int.\ Jour.\ Theor.\
	Phys.} {\bf 31} (1992) 789--807.

\bibitem{Pul95} S.~Pulmannov\'a, ``Difference posets and the
	histories approach to quantum theories'' {\em Int.\ Jour.\
	Theor.\ Phys.} {\bf 34} (1995) 189--210.

\bibitem{BLM91} P.~Busch, P.~J.~Lahti and P.~Mittelstaedt, {\em The
	Quantum Theory of Measurement}, (Springer-Verlag, Berlin, 1991).

\bibitem{FGR93} D.~J.~Foulis, R.~J.~Greechie and G.~T.~R\"uttimann,
	``Logicoalgebraic structures II. Supports in test spaces'' {\em
	Int.\ Jour.\ Theor.\ Phys.} {\bf 32} (1993) 1675--1690.

\bibitem{HF81} G.~M.~Hardegree and P.~J.~Frazer, ``Charting the
	labyrinth of quantum logics: A progress report'' in {\em Current
	Issues in Quantum Logic\/}, eds E.~G.~Beltrametti and B.~V.~van
	Frassen (Plenum Press, London 1982).

\bibitem{Gri95} R.~Griffiths  ``Consistent quantum reasoning'',
	Carnegie--Mellon University preprint (1995).

\bibitem{Gle57} A.~M.~Gleason, ``Measures on the closed subspaces of
	a Hilbert space'' {\em Jour.\ Math.\ Mech.} {\bf 6} (1957) 885--893.

\bibitem{ILS94} C.~J.~Isham, N.~Linden and S.~Schrekenberg, ``The
	classification of decoherence functionals: An analogue of Gleason's
	theorem'' {\em 	Jour.\ Math.\ Phys.} {\bf 35} (1994) 6360--6370.

\bibitem{Sch95} S.~Schreckenberg, ``Completeness of Decoherence
	Functionals'', to appear {\em Jour.\ Math.\ Phys.} (1995).

\bibitem{Wri95} J.~D.~M.~Wright, ``The structure of decoherence
	functionals in von~Neumann quantum histories'', to appear {\em
	Jour.\ Math.\ Phys.} (1995).

\bibitem{BMW94} L.~J.~Bunce and J.~D.~M.~Wright, ``The
	Mackey--Gleason problem for vector measures on projections in
	von Neumann algebras'' {\em Jour.\ Lond.\ Math.\ Soc.} {\bf 49}
	(1994) 133--149.

\bibitem{Ish84} C.~J.~Isham, ``Topological and global aspects of
	quantum theory'' in {\em Relativity, Groups and Topology II},
	eds C.~DeWitt and R.~Stora, (North-Holland, Amsterdam, 1984).

\bibitem{IL95} C.~J.~Isham and N.~Linden, ``Continuous histories
	and the history group in generalised quantum theory'',
	to appear {\em Jour.\ Math.\ Phys.} (1995).

\bibitem{Gui72} A.~Guichardet, {\em Symmetric Hilbert Spaces and
	Related Topics; Springer Lecture Notes in Mathematics, Vol.~261},
	(Springer-Verlag, Berlin, 1972).
\end{thebibliography}
\end{document}